# The geometry of high-dimensional phase diagrams III: Engineering relative stability in four dimensions

Jiadong Chen[1], Matthew J. Powell-Palm,[2,3] Wenhao Sun[1*]
[1]Department of Materials Science and Engineering, University of Michigan, Ann Arbor, MI, 48109, USA
[2]J. Mike Walker '66 Department of Mechanical Engineering, Texas A&M University, College Station, TX, 77802, USA
[3]Department of Materials Science & Engineering, Texas A&M University, College Station, TX, 77802, USA

*Correspondence to: whsun@umich.edu

## Significance Statement

Designing thermodynamic conditions to improve (or reduce) the stability of a target material is a key task in materials engineering. For example, during materials synthesis one aims to enhance the stability of a target phase relative to its precursors or competing byproduct phases. If an undesired phase forms in experiment, one aims to destabilize the undesired phase by dissolution or corrosion. When multiple thermodynamic knobs are available to engineer relative stability, it can be difficult to navigate the corresponding high-dimensional phase diagram to identify optimal pathways to promote or destabilize a target phase. We propose that instead of mapping the absolute phase boundaries of a target material, we can invoke a generalized Clausius-Clapeyron relation, which provides a 'compass' to point out which directions on a high-dimensional phase diagram are best to stabilize or destabilize a target phase.

## Abstract

Phase boundaries on high-dimensional phase diagrams are also high-dimensional objects, and can represent phase coexistence between numerous phases simultaneously. On a 2D temperature-pressure phase diagram, phase boundaries are 1D lines separating two phases, with a slope given by the Clausius-Clapeyron relation, $dP/dT = \Delta S/\Delta V$. However, this derivative form of the Clausius-Clapeyron relation does not scale meaningfully to higher dimensions. Here, we derive a parametric form of the Clausius-Clapeyron relation that is readily applicable to high-dimensional phase diagrams. The gradient of this phase boundary guides us on how to increase or decrease the relative stability of a target compound, meaning this generalized Clausius-Clapeyron relation enables us to engineer relative stability with respect to multiple thermodynamic conditions simultaneously. Using this approach, we analyze the acid stability of manganese oxide catalysts on a 4-dimensional Pourbaix diagram with axes of $p$H, redox potential, nanoparticle size, and aqueous $[K^+]$ ion concentration.

A primary goal in materials thermodynamics is to construct phase diagrams with accurate and precise phase boundaries for all known phases in a chemical system. However, the 'thermodynamic assessments' required to construct accurate phase diagrams[1,2,3,4] can be very time-consuming. This process involves compiling all calorimetric and DFT-computed thermochemical data, constructing free-energy models for the solids, solid-solutions and liquid phases, then critically evaluating the resulting phase diagrams against experimentally-observed phase boundaries, while adhering to Gibbs Phase Rule and other thermodynamic considerations of phase coexistence.[5] Because thermodynamic assessments can be such a laborious process, phase diagrams do not exist in many materials engineering contexts, despite their obvious importance and utility.

It would be valuable to develop a simpler and more agile framework to explore the stability conditions of materials. We propose that in many materials engineering situations, it may not actually be necessary to map out the full phase diagram. Instead, it may be enough to just characterize the stability of a desired target phase *relative* to its competing phases. For example, perhaps we aim to synthesize a target phase $\alpha$, but experimentally we observe that $\alpha$ transforms to an undesired $\beta$ phase, or that formation of $\alpha$ is blocked by the undesired nucleation of $\gamma$, *etc*. The salient question then becomes: How do we modify our experimental conditions to promote the stability of $\alpha$, while destabilizing all other competing phases? In other words, how do I engineer the *relative stability* between a desired target phase versus an experimentally-obtained undesired phase?

High-dimensional phase diagrams become increasingly difficult to navigate when there are multiple operative forms of thermodynamic work being considered. To engineer stability in multi-parameter thermodynamic space, materials scientists have often turned to statistical approaches, such as Design of Experiments (DoE), Bayesian optimization with Gaussian processes, or other recent AI approaches in sequential learning.[6,7,8,9,10,11] These sequential learning algorithms are usually physics-agnostic, however, their efficiency and efficacy can be greatly improved by augmenting them with stronger thermodynamic priors on how to engineer relative stability.[12]

When considering relative stability, the most important feature of a thermodynamic phase diagram is the gradient of the phase boundary between the target phase and its competing phase(s). This is formulated in the Clausius-Clapeyron relation, which on a 2D temperature-pressure diagram is derived from $dG_\alpha = dG_\beta$; such that$-S_\alpha dT + V_\alpha dP = -S_\beta dT + V_\beta dP$; resulting in $dP/dT = \Delta S / \Delta V$. As shown in **Figure 1**, isothermally pressurizing a fluid usually enhances the relative stability of the solid—as $dP/dT > 0$, and $\Delta S_{S\rightarrow L} > 0$ while usually $\Delta V_{S\rightarrow L} > 0$. However, isothermally pressurizing $H_2O$ preferences the stability of water over ice, as $\Delta V_{I\rightarrow W} < 0$, as water is denser than ice.

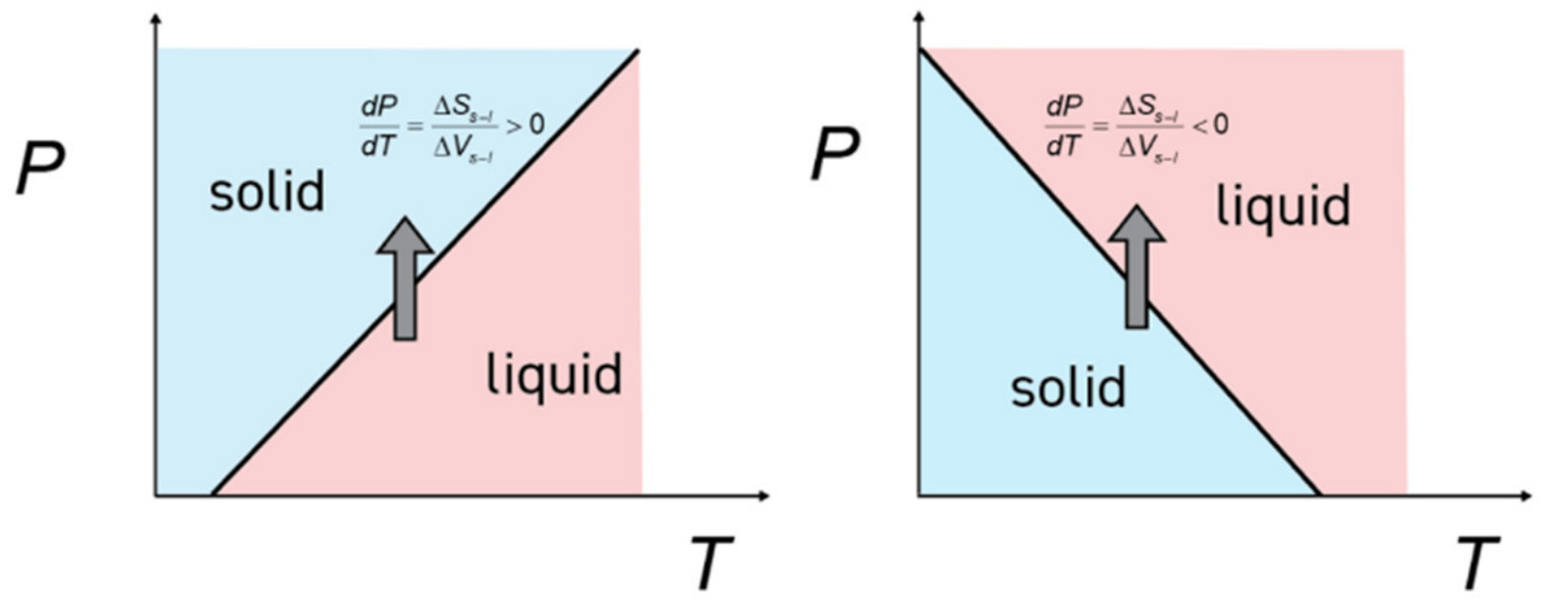

**Figure 1.** The gradient of a phase boundary determines the relative stability between two or more phases under changing thermodynamic conditions. **Left** shows an example where the solid is denser than liquid; whereas **Right** shows then liquid is denser than solid (like $H_2O$).

Beyond temperature and pressure considerations, a generalized Clausius-Clapeyron relation can be derived from a thermodynamic potential $dZ = -X_1 dY_1 - X_2 dY_2$, where $Y_1$ and $Y_2$ are intensive thermodynamic variables, which yields the equation $dY_1/dY_2 = -\Delta X_2/\Delta X_1$. Generalized Clausius-Clapeyron relations have been used to determine $dT/dH$ for magnetic materials,[13] $dF/dT$ for shape-memory alloys,[14] and even $d(pH)/d\sigma$ for martensitic actuators in viruses.[15] The derivative form of the Clausius-Clapeyron equation also serves as a starting point to more complicated thermodynamic relations—for example by combining with Maxwell's relations or Bridgman's relations[16] to derive new thermodynamic partial derivatives for a variety of materials engineering applications.

Unfortunately, the derivative form of the Clausius-Clapeyron equation $dY_1/dY_2$ is not readily generalizable to higher-dimensional phase diagrams, as the phase-coexistence regions become higher-dimensional than a 1D line. For example, on a 4-dimensional phase diagram, one can have phase boundaries that are 0-, 1-, 2-, or 3-dimensional; corresponding to 5-, 4-, 3-, or 2-phase coexistence, respectively. For higher dimensional phase boundaries, it is not meaningful to write derivative-like expressions for gradients between three or more variables (such as dT/dP/dH/…/*etc.*); nor is it straightforward to arrive at such a ratio starting from $dG_\alpha = dG_\beta = dG_\gamma = \ldots$ *etc.*

Here, we derive a vector representation of the Clausius-Clapeyron relation, which is readily generalizable to high-dimensional phase boundaries. We develop two representations of this relation: 1) a parametric representation for "bottom-up" construction of high-dimensional phase boundaries, where the phase boundaries are constructed from known or measured extensive variables of the competing phases; and 2) a Cartesian representation for "top-down" half-space intersections of high-dimensional free energy surfaces. Note that Clausius-Clapeyron analyses *only* work on phase diagrams with all-intensive variables, which can be generally derived using the duality approach discussed in Part II of this three-part series. Because we are concerned with only the stability of a single target phase, phase diagrams with intensive natural variables are ideal to engineer relative stability while avoiding the complexities of heterogeneous equilibrium.

First, we demonstrate the bottom-up approach to the high-dimensional Clausius-Clapeyron relation by calculating the 2D phase boundary on a 3D temperature-pressure-magnetic field (*T*-*P*-*H*) phase diagram between BCC and FCC iron, using their known entropies, molar volumes, and magnetic moments. A major advantage of the Clausius-Clapeyron equation is that it does not require free energies, it only needs the extensive quantities of the two phases, which can often be easier to obtain than the temperature- and pressure-dependent Gibbs free energies of each phase. By knowing the high-dimensional gradient of the phase boundary, along with a single multi-phase coexistence point, the entire phase boundary can be constructed parametrically.

Next, we demonstrate the 'top-down' approach by calculating, visualizing, and interpreting a full 4D phase diagram. Specifically, we build a 4D Pourbaix diagram to examine the acid stability of manganese oxides, which have applications as earth-abundant oxygen evolution catalysts.[17,18] We extend the traditional *p*H and *E* axes in two additional dimensions to further account for nanoparticle size (1/R) and impurity ion concentration ($\mu_{K^+}$). We focus our discussion on how to leverage the generalized Clausius-Clapeyron relation to navigate non-intuitive aspects of high-dimensional phase diagrams, from which we can derive meaningful insights to engineer relative materials stability.

## Vector derivation of the Clausius-Clapeyron Relation

We begin by rederiving the classic 2D Clausius-Clapeyron relation from a vector representation. On a *T-P* phase diagram, the Gibbs free-energy surface of a phase is $G = H' + PV - TS$; where $H'$ is the standard-state formation enthalpy of a phase, and $P' = P - 1$ atm (since $P = 1$ atm at standard state). This can be rewritten in Cartesian form such that $ax + by + cz - d = 0$ has a 1-to-1 mapping to the expression $ST - VP' + G - H' = 0$. In our goal of calculating gradients, we treat *S* and *V* as constant at a given *T* and *P*, which linearizes the free-energy plane in the vicinity of a given phase coexistence point.

On the *T–P–G* axes, the normal vector of the free-energy plane can be expressed as $\langle S_i, -V_i, 1\rangle$ where *i* represents the phase. **Figure 2** illustrates an example of the ice/water phase boundary at 273K and 2.16 MPa, with a table showing their corresponding entropies and molar volumes.[19] Between two phases, α and β, the normal vectors of their free-energy planes are $n_\alpha$ and $n_\beta$, and their cross product $n_\alpha \times n_\beta$ produces the differential vector for the phase coexistence line in *T–P–G* space:

$$\begin{vmatrix} \hat{T} & \hat{P} & \hat{G} \\ S_\alpha & -V_\alpha & 1 \\ S_\beta & -V_\beta & 1 \end{vmatrix} = \langle V_\beta - V_\alpha,\ S_\beta - S_\alpha,\ V_\alpha S_\beta - S_\alpha V_\beta \rangle$$

where $\hat{T}, \hat{P}$, and $\hat{G}$ are unit vectors in the temperature, pressure and Gibbs free energy direction with their appropriate corresponding units. The parametric form of the 1D phase-coexistence line, *L*, in *G*(*T*,*P*) space can therefore be expressed as:

$$L_{Coexistence} = \langle T_0, P_0, G_0 \rangle + \langle \Delta V,\ \Delta S,\ V_\alpha S_\beta - S_\alpha V_\beta \rangle \lambda$$

where λ is the parameter, $\langle T_0, P_0, G_0 \rangle$ is an initial condition point. This form also produces an expression for $dG = V_\alpha S_\beta - S_\alpha V_\beta$ along the coexistence line, which does not appear in the traditional Clausius-Clapeyron relation.

By projecting this vector onto the *T–P* axes (in other words, eliminating the G term) and expressing this coexistence line in a parametric differential form, we recover the classical Clausius-Clapeyron relation, $dT/dP = \Delta V/\Delta S$, rewritten in differential vector form as $\langle dT, dP \rangle = \langle \Delta V, \Delta S \rangle_{T,P} d\lambda$ where $\lambda$ is the parameter. Importantly, to preserve the units of the *T*, *P* and *G* axes, λ must have units of [Temperature / Volume], which ensures for example that $dT$ and $\lambda\Delta V$ both have units of temperature. More generally, the parameter λ has units of [Energy / ([$X_1$][$X_2$])], where [$X_1$] and [$X_2$] are the units of the conjugate extensive variables $X_1$ and $X_2$.

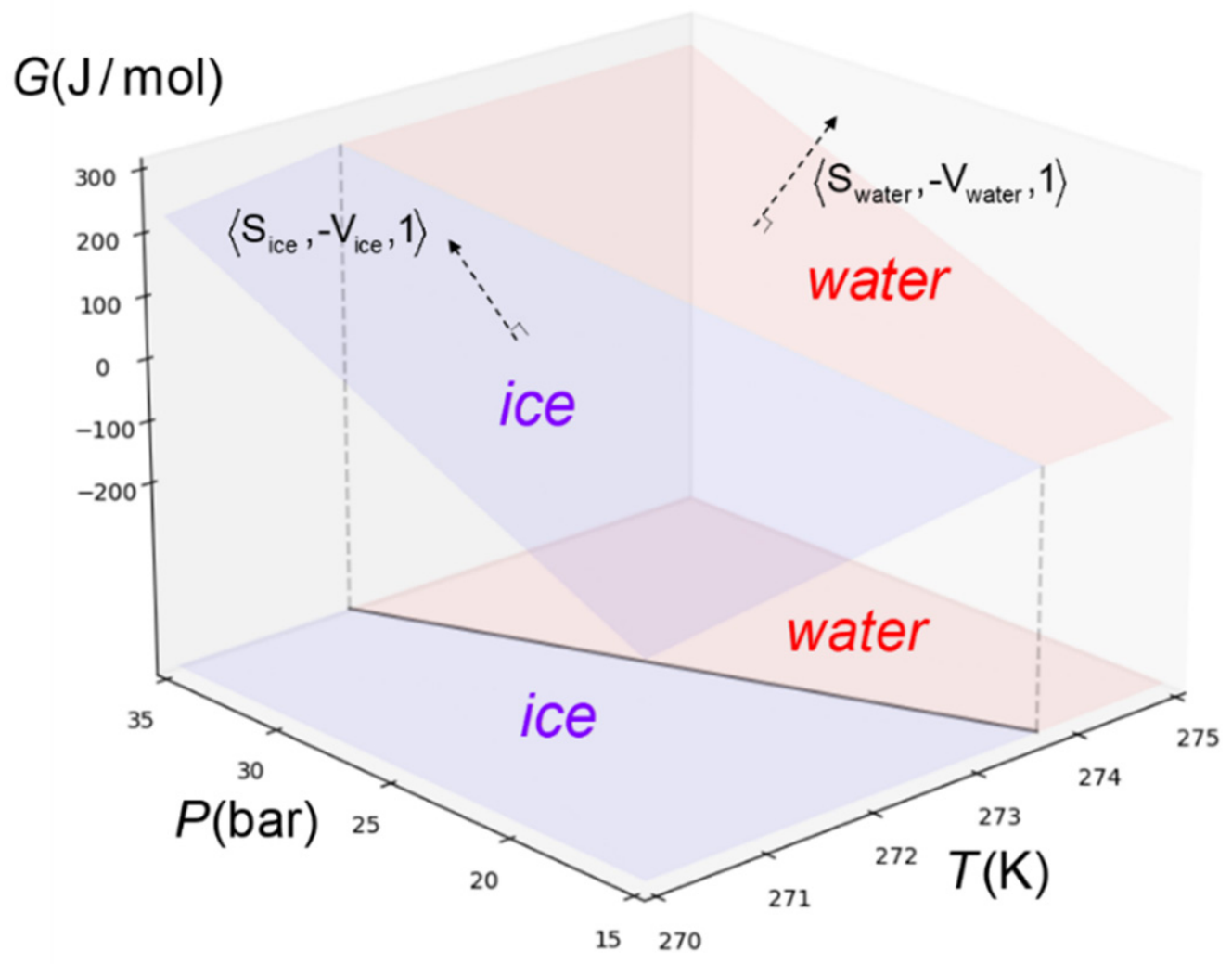

| | $T_0$ | $P_0$ |
|---|---|---|
| **Condition** | 273 K | 2.16 MPa |
| | **S** (J/mol/K) | **V** ($cm^3$/mol) |
| **Water** | -2.558 | 17.19 |
| **Ice** | -23.55 | 19.39 |
| $\Delta X_{\text{Water-Ice}}$ | 20.99 | -2.2 |

**Figure 2. Vector representation of Clausius-Clapeyron relation for the ice-water coexistence boundary**. The coexistence vector is given by the cross product of the normal vectors to each free-energy plane in $G(T,P)$ space. **Table 1.** Molar entropies and volumes between water and ice at a given $T_0$, $P_0$ condition.

To construct a phase diagram, one generally requires the free-energies of all phases in a chemical system. However, obtaining free-energies may not always be possible or convenient. In a 'bottom-up' approach to the Clausius-Clapeyron relation, one can parametrically construct the entire phase boundary with just the extensive variables for all competing phases, plus one point of coexistence to anchor the phase boundary. The gradient of a phase boundary can be linearized by assuming the extensive variables are constant—this will apply for small perturbations in intensive conditions, but for curved phase boundaries one should recalculate the $\Delta X$ at other coexistence points.

## High-Dimensional Generalization of the Clausius-Clapeyron Relation

On a high-dimensional phase diagram with all intensive axes, phase boundaries can be up to $k$-dimensional for any integer $k < d$, where $d$ is the dimensionality of the phase diagram. This $k$-dimensional phase boundary represents phase coexistence between $(d - k)$ phases, and can be spanned by a linear combination of $k$ one-dimensional Clausius-Clapeyron vectors, built from any two intensive variables.

Consider the allotropic phase transformation between BCC and FCC iron. In addition to the temperature and pressure driven phases transformations between these two phases, BCC iron also has a higher magnetic moment than FCC Fe (2.2 $\mu_B$ vs 1.5 $\mu_B$), so their phase boundary should also vary with the applied magnetic field. The thermodynamic potential for a single-component material with temperature, pressure, and magnetic field as natural variables is $dZ = -SdT + VdP - MdH$. This is a 4-dimensional free-energy space (1 energy axis and 3 work axes). By Generalized Gibbs' Phase Rule[20], $F = W - P + 1$. With 3 thermodynamic axes ($W = 3$) and two-phase coexistence ($P = 2$), the phase boundary between BCC and FCC iron has 2 intensive degrees-of-freedom ($F = 2$), meaning it is a 2-dimensional surface on the 3D $T$-$P$-$H$ phase diagram.

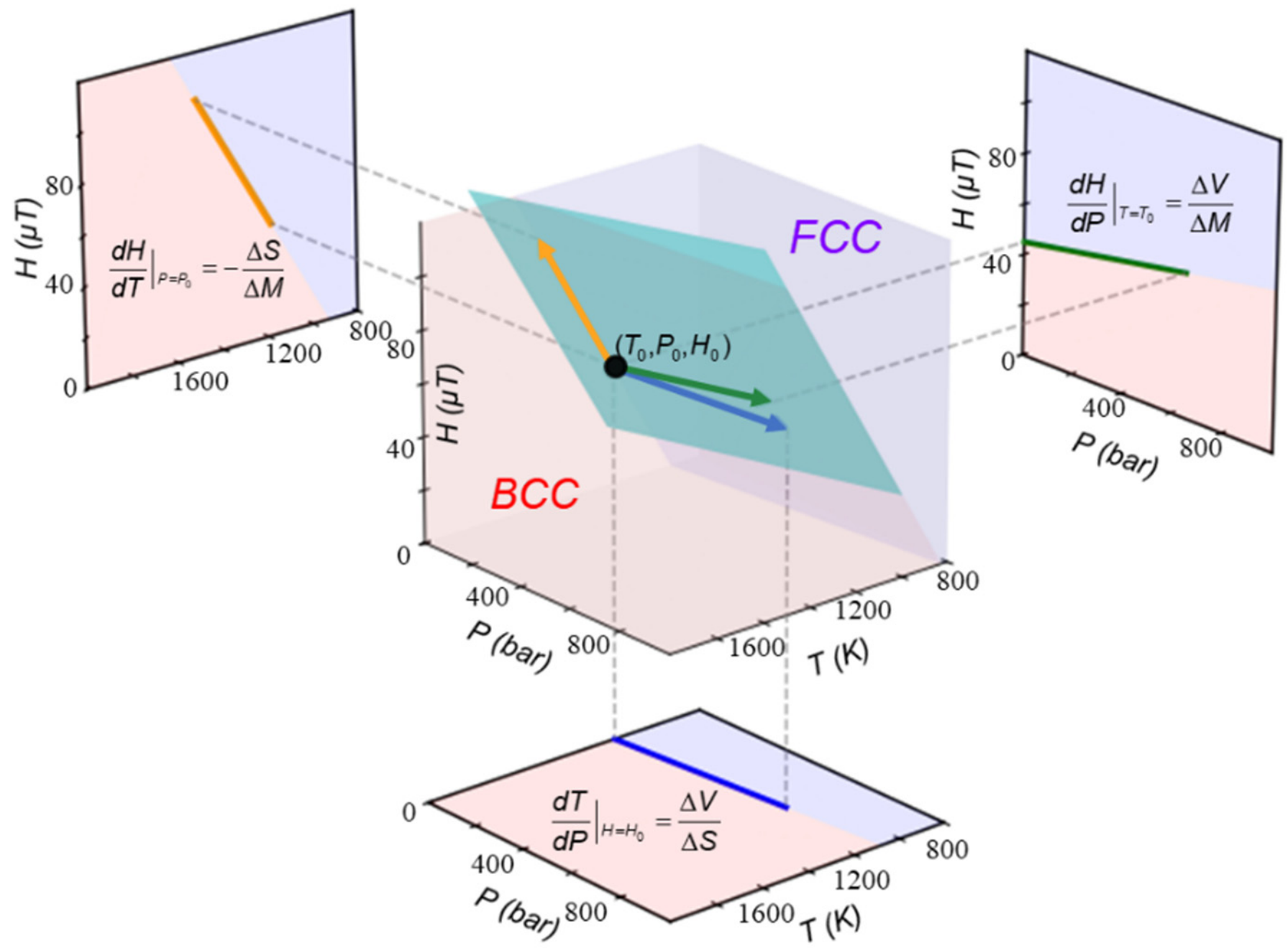

| | $T_0$ | $P_0$ | $H_0$ |
|---|---|---|---|
| **Condition** | 1183 K | 1 atm | 0.45 μT |
| | **S** (J/mol/K) | **V** ($cm^3$/mol) | **M** ($\mu_B$) |
| **α-Fe (BCC)** | 27.28 | 7.37 | 2.2 |
| **γ-Fe (FCC)** | 27.97 | 7.30 | 1.5 |
| **$\Delta X_{FCC\text{-}BCC}$** | 0.69 | -0.07 | -0.7 |

**Figure 3. Clausius-Clapeyron relation on 3D phase diagram of temperature-pressure-magnetic field**, showing a 2D phase boundary between ferromagnetic *α*-Fe (BCC) and paramagnetic *γ*-Fe (FCC). The gradient of this 2D phase boundary is constructed parametrically, taken from the linear combination of 2D Clausius-Clapeyron relations between *H*-*T*, *T*-*P*, or *H*-*P* axes. Blue is *α*-Fe (BCC) and red is *γ*-Fe (FCC). **Table 2.** Molar entropies, volumes, and magnetic moments between BCC and FCC iron at a given $T_0$, $P_0$, $H_0$ condition.[21,22]

**Figure 3** illustrates the 2D phase boundary between FCC and BCC iron on the 3D *T-P-H* phase diagram. There are three possible 1D vectors for the 2D Clausius-Clapeyron relation:

$$\left\langle dT, dP, dZ \right\rangle_{dH=0} = \left\langle \Delta V, \Delta S, V_\alpha S_\beta - S_\alpha V_\beta \right\rangle_{T,P,H} d\lambda$$

$$\left\langle dT, dH, dZ \right\rangle_{dP=0} = \left\langle -\Delta M, \Delta S, S_\alpha M_\beta - M_\alpha S_\beta \right\rangle_{T,P,H} d\mu$$

$$\left\langle dP, dH, dZ \right\rangle_{dT=0} = \left\langle -\Delta M, -\Delta V, M_\alpha V_\beta - M_\beta V_\alpha \right\rangle_{T,P,H} d\nu$$

Where the units of the parameters are $\lambda \equiv$ [Temperature / Volume], $\mu \equiv$ [Temperature / Magnetic Moment], and $\nu \equiv$ [Energy / Volume / Magnetic Moment]. The 2-dimensional boundary between BCC and FCC Fe can be spanned by a linear combination of any two of these 1D vectors within this 4D space, such that the plane is expressed $P = P_0 + \alpha \mathbf{v}_1 + \beta \mathbf{v}_2$, where α and β are arbitrary parametric terms.

We can further sum all three vectors together and normalize, resulting in a symmetric form of:

$$\left\langle dT, dP, dH \right\rangle = \frac{1}{2} \left\langle \Delta V\, d\lambda - \Delta M\, d\mu,\ \Delta S\, d\lambda - \Delta M\, d\nu,\ \Delta S\, d\mu - \Delta V\, d\nu \right\rangle$$

Finally, although the original Clausius-Clapeyron relation does not give the change in free energy along the coexistence boundary, we can further calculate it within our representation:

$$dZ = d\lambda (V_\alpha S_\beta - S_\alpha V_\beta)_{dH=0} + d\mu (S_\alpha M_\beta - M_\alpha S_\beta)_{dP=0} + d\nu (M_\alpha V_\beta - M_\beta V_\alpha)_{dT=0}$$

In even higher dimensions, one can continue this process simply by taking additional linear combinations of 1D Clausius-Clapeyron vectors. As shown in the *T-P-H* phase diagram, only 2 of 3 possible Clausius-Clapeyron vectors are needed to span the 2D phase boundary. In higher dimensions, the number of required 1D vectors diminishes combinatorically. For any *k*-dimensional phase boundary, *k* Clausius-Clapeyron vectors are required to span the phase boundary, but the number of possible $\partial Y_1/\partial Y_2$ relations scales as ${}_dC_2$. This enables one to infer missing Clausius Clapeyron relations and their corresponding extensive materials properties, which is especially valuable when some partial derivatives or properties are difficult to measure or calculate. A full discussion of this scaling relationship is in **Supplemental Information 1**.

**Cartesian representation of Clausius-Clapeyron relationships**

Thus far we have described a parametric vector approach to building high-dimensional phase boundaries, which are spanned by linear combinations of 1-dimensional Clausius-Clapeyron vectors. In this approach, we need as many 1D vectors as there are dimensions of the phase boundary. This approach is most applicable in experimental contexts with limited materials properties or phase equilibria data, or when rigorous free-energy descriptions of the relevant phases are unavailable. However, in the era of high-throughput computational materials science, the free energy surfaces of phases can be calculated computationally. The opportunity arises then to leverage these free energies to better understand phase equilibria in complex thermodynamic environments.

To this end, we next present an alternative "top-down" approach to analyze high-dimensional phase boundaries, which assumes the availability of free-energy data. This approach, which arrives at mathematically-equivalent descriptions of the phase boundary compared to the vector approach, identifies high-dimensional phase boundaries by calculating the half-space intersections of high-dimensional free-energy hyperplanes.

In an $N$-dimensional thermodynamic space (including the energy axis), the free-energy, $Z$, of a single phase can be represented in Cartesian form by

$$\sum_{i,d} X_i Y_i - Z = 0$$

In this case, a $k$-dimensional phase boundary can be calculated from the intersection of ($N$–$k$) hyperplanes from the $N$-dimensional thermodynamic space. For example, in a three-dimensional $G(T,P)$ space, the free-energy planes are two-dimensional, and a two-phase coexistence 1D line is given by the intersection of 2 hyperplanes. A three-phase coexistence 0D point is given by the intersection of 3 hyperplanes. This argument extends to higher-dimensions; e.g. two-phase coexistence is given by the intersection of two 4D hyperplanes, which results in a 3-dimensional phase boundary.

One can express these intersections by equating the hyperplane equations, for example, on a 4D phase diagram, three-phase coexistence between the phases $\alpha, \beta, \gamma$ is given on a 2D phase boundary, which can be written as $Z_\alpha = Z_\beta = Z_\gamma$:

$$\sum_i X_{i,\alpha} Y_i = \sum_i X_{i,\beta} Y_i = \sum_i X_{i,\gamma} Y_i$$

One major benefit of this Cartesian representation is that the Clausius-Clapeyron relationship for two-phase coexistence becomes simple to compute in high dimensions. In parametric form, a $k$-dimensional phase boundary needs to be spanned by $k$ one-dimensional vectors, meaning for two-phase coexistence one needs ($d$–1) individual 1D Clausius-Clapeyron vectors. However, if one has the free-energy surfaces, one can directly compute any Clausius-Clapeyron derivative simply by calculating $dY_1/dY_2 = -\Delta X_2/\Delta X_1$, where the extensive variables $X$ are parameters in the Cartesian free-energy expression.

## Clausius-Clapeyron Analysis of a 4D Pourbaix Diagram.

We conclude this three-part series on high-dimensional thermodynamics with a Clausius-Clapeyron analysis of a full 4-dimensional phase diagram. Our goal is to illustrate how the simultaneous consideration of multiple thermodynamic variables enables a more comprehensive approach to materials design and engineering. Although a 4D phase diagram stretches our imagination, it is still conceptually accessible by imagining the fourth dimension as time. While examining this 4D diagram, we will develop tools and intuition to mathematically conduct dimensional analogy to even higher-dimensional phase diagrams.

In particular, here we analyze the stability of manganese oxides under acidic conditions, with a goal to increase the stability of a solid manganese oxide relative to its dissolved $Mn^{2+}$ aqueous ion. Energy storage and transformation technologies require new catalysts[23,24] for the oxygen reduction reaction (ORR) and oxygen evolution reaction (OER) [25,26], ideally without using expensive noble metal catalysts like platinum[27,28,29]. One candidate system is manganese oxide-based catalysts, however manganese oxides are generally not stable in acid electrochemical environments, where they easily decompose during changes in redox potential during cyclic voltammetry.[30,31,32]

There is a great diversity of manganese-based oxide materials, with various polymorphs and manganese oxidation states.[33,34,35] This raises the question of whether or not there exists a candidate manganese oxide phase that has good stability in acidic solutions. As illustrated on a Pourbaix diagram, the relative stabilities of different manganese oxides vary as a function of aqueous *p*H and *E*. Furthermore, different surface energies between manganese oxide phases can drive nanoscale crossovers in polymorph stability,[33,36,37] and intercalation of impurity ions from solution such as $K^+$, $Na^+$, $Ca^{2+}$, *etc,* can also affect the bulk stability of various polymorphs.[33]

To capture all of these effects simultaneously, we construct here a four-dimensional Pourbaix diagram for the Mn-$H_2O$ system with axes of *p*H, redox potential, nanoparticle size, and [$K^+$] impurity ion concentration. The composition- and size-dependent Pourbaix potential for each phase can be written as:

$$\Psi = \frac{1}{N_m}\left( (G_{bulk} - N_O\mu_{H_2O}) - RT\ln(10)(2N_O - N_H)p\mathrm{H} - (2N_O - N_H + Q)E + (\frac{A}{V})\gamma\eta\rho - N_K\mu_K \right)$$

The derivation of this potential is provided in our previous works.[36,37] Here, $\Psi$ is the Pourbaix potential, with respect to *p*H; redox potential, E; surface area to volume ratio, $A/V$; chemical potential of potassium, $\mu_K$, under a constraint of water-oxygen equilibrium. $N$, is the number of atoms of a certain element; $Q$, is the number of charges; $\rho$, is volume density; $\eta$, is shape factor; $\mu_{H2O}$, is water energy; $\mu_K$, is surface energy. The number of Mn atoms are conserved in the phase transformations between Mn-based oxides with different compositions, thus $\Psi$ is normalized by the number of Mn atoms, $N_m$. The molar Gibbs free energy of a phase, $G_{bulk}$, is its chemical potential, $\mu_i = \mu_i^o + RT\ln[a_i]$, where $\mu_i^o$ is given by the standard-state Gibbs formation free-energy, $\Delta G_f^o$, and the activity ideally scales with the natural log of the metal ion concentration in solution.

In our analysis, we consider the phases: $Mn^{2+}$, $MnO_4^-$, $Mn_3O_4$, α-$Mn_2O_3$, *α*-MnOOH, *γ*-MnOOH, *α*-$K_{0.11}MnO_{1.94}$, *δ*-$K_{0.21}MnO_{1.87}$, *γ*-$MnO_2$ and *β*-$MnO_2$. The thermochemical data for these phases, including surface energies and $K^+$-intercalated energies, were calculated in our previous publications from DFT using the SCAN metaGGA functional.[38,39,40] Our thermochemical data used here is tabulated in the **Supplementary Table 2**.

Because we are investigating the acid stability of solid manganese oxides, the relevant phase boundary is between each solid manganese oxide phase and its dissolved state, the $Mn^{2+}$(aq) ion. From a Clausius-Clapeyron perspective, we aim to increase the relative stability of the solid manganese oxide, meaning we need to determine how varying these 4 thermodynamic variables will shift the phase boundary of a manganese oxide solid into and towards the $Mn^{2+}$(aq) region, thereby enlarging the stability region of the solid. The relative stability analysis is then to find conditions where the Clausius-Clapeyron relation $d\mathit{p}H/dY < 0$; such that a change in $Y$ shifts the phase boundary between $Mn^{2+}$(aq) vs. $MnO_x$(solid) to lower pH values, indicating increased acid stability.

## Phase Coexistence on a High-Dimensional Phase Diagram

All the variables in the size-dependent Pourbaix potential are intensive, meaning that phase stability regions are all 4-dimensional stability polytopes—in other words, single-phase regions all have four intensive degrees of freedom. Phase coexistence boundaries all have $F = 5 - P$ degrees of freedom, where $F$ is the dimensionality of the phase boundary and $P$ is the number of coexisting phases, as summarized in **Table 3.** Some non-intuitive aspects of high-dimensional geometry emerge—for example, it is possible for three 4D single-phase regions to coexist on a 2D phase boundary; also, the phase boundary between two phases is 3-dimensional. These facts are difficult to visualize in our three-dimensional universe, but they are direct consequences of generalized Gibbs' phase rule.

**Table 3**. Formulas of Coexisting Phases and their Dimensionality of Phase Boundary.

| *P*-phase coexistence | Dimensionality of Phase Boundary | Coexisting Phases |
|---|---|---|
| 5 | 0 (vertex) | *β*-$MnO_2$, *α*-$K_{0.11}MnO_{1.94}$, *δ*-$K_{0.21}MnO_{1.87}$, $Mn^{2+}$, *α*-MnOOH |
| 4 | 1 (line) | *β*-$MnO_2$, *α*-$K_{0.11}MnO_{1.94}$, *δ*-$K_{0.21}MnO_{1.87}$, $Mn^{2+}$ |
| 3 | 2 (polygon) | *α*-$K_{0.11}MnO_{1.94}$, *δ*-$K_{0.21}MnO_{1.87}$, $Mn^{2+}$ |
| 2 | 3 (polytope) | *δ*-$K_{0.21}MnO_{1.87}$, $Mn^{2+}$ |
| 1 | 4 (polytope) | *δ*-$K_{0.21}MnO_{1.87}$ |

To facilitate the visualization of these 4D phase stability regions and their corresponding phase boundaries, here we introduce the concepts of *slice* and *projection*, illustrated in **Figure 4**, which are two different approaches to dimensionality reduction for 2D or 3D visualization. To make a slice, an intensive variable is set to a constant value and the rest of the thermodynamic potential is evaluated. Slices essentially remove one dimension from the phase diagram. The other dimensionality reduction method is *projection*, which shows a 'shadow' of the phase on the thermodynamic axes, constructed by projecting all the vertices of the stability region onto the axes, and taking the geometric convex hull of the vertices. One limitation of the projection approach is that when projecting multiple phases, the 'shadows' of various phases can overlap. However, the advantage of the projection is that it reveals all the possible thermodynamic conditions a phase can exist; whereas one would typically have to construct slices sequentially over a thermodynamic axis to survey all the possible stability conditions of a phase. Once the domain of stability for a desired phase is determined from a projection approach, one can further apply slices to study materials stability to construct interpretable phase diagrams.

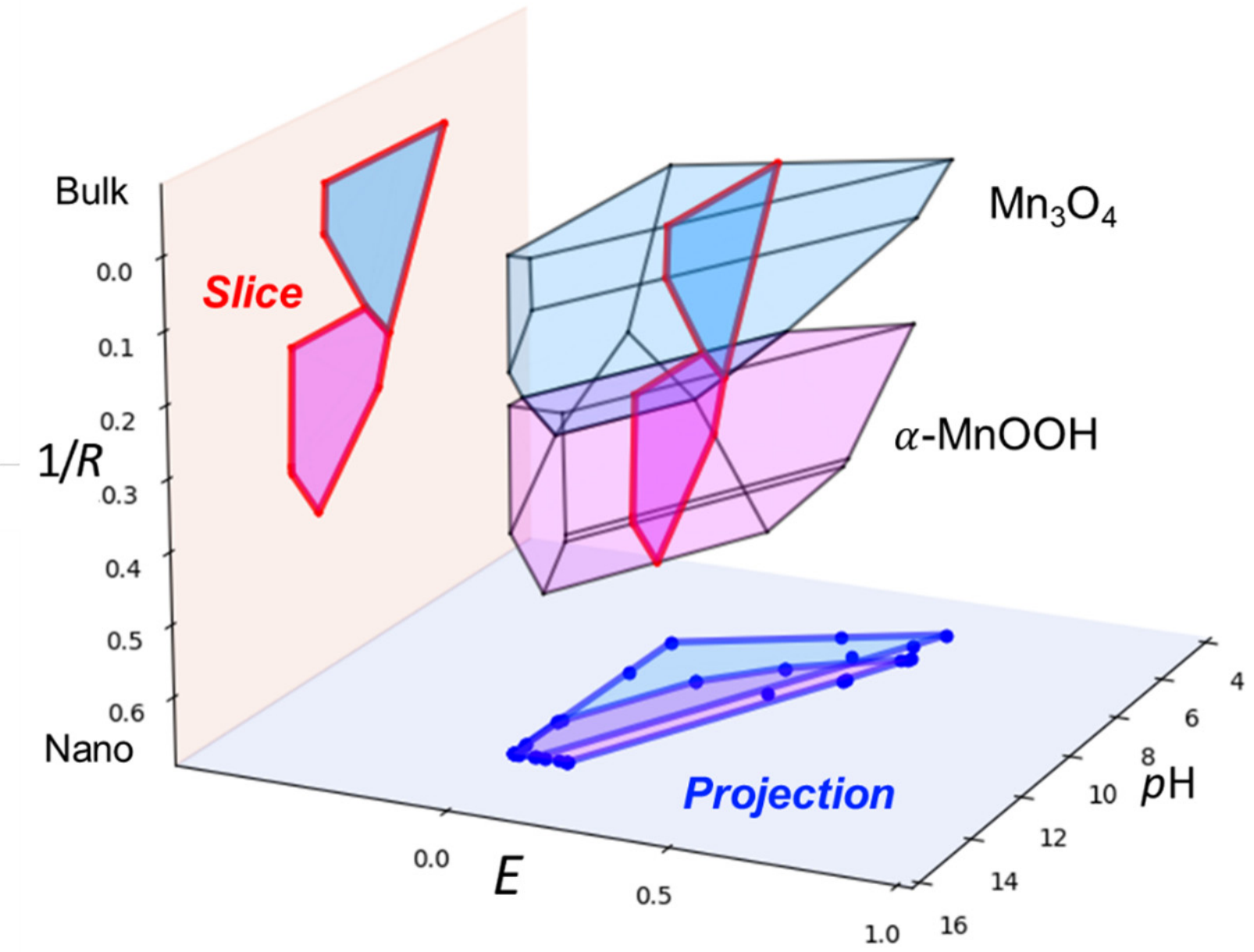

**Figure 4**. **Phase stability region of $Mn_3O_4$ in 1/*R*, *E*, *p*H space.** The plane bounded by red lines is a slice when E is fixed at 0.2V. The plane bounded by blue lines is a projection to *p*H -*E* space.

## 4D size- and impurity concentration-dependent Pourbaix Diagram

**Figure 5** shows several different perspectives of our 4-dimensional Pourbaix diagram. Because we want to improve the acid stability of manganese oxides, we focus our diagrams on the solid manganese oxide phases that border the $Mn^{2+}$(aq) ion, which is the undesired dissolution product of solid manganese oxides in acid. **Figure 5a** shows the traditional 2D Pourbaix diagram for the Mn-$H_2O$ system, which visualizes the bulk equilibrium phases under a given *E* and *p*H. The bulk equilibrium phase *β*-$MnO_2$ is only stable at low *p*H in a small range of high redox potentials.

For catalysts, it is often valuable to maximize the surface area to volume ratio. This provides the greatest amount of active catalytic area for a given mass of catalytic material. Not only does a high surface area to volume ratio promote the functional performance of a catalyst, a variety of metastable manganese oxide phases can be stabilized at high surface-area-to-volume ratios—as demonstrated in previous experimental investigations[41,42,43,44], as well as our previous computational studies.[36] These metastable manganese oxides phases have lower surface energy than the bulk equilibrium phases, so at the nanoscale where surface-area-to-volume ratio is large, these bulk metastable phases can in fact become the nanoscale equilibrium phase.

In our previous work, we visualized nanoscale Pourbaix diagrams[36] using 2D slices of the (1/R) axis at fixed *E* or fixed *p*H. In this work, **Figure 5b** shows the full 3D nanoscale Pourbaix diagram, along with the competing nanoscale crossovers in polymorph stability. Size-stabilized bulk metastable manganese oxides include α-MnOOH, γ-MnOOH, δ-$MnO_2$ and R-$MnO_2$.

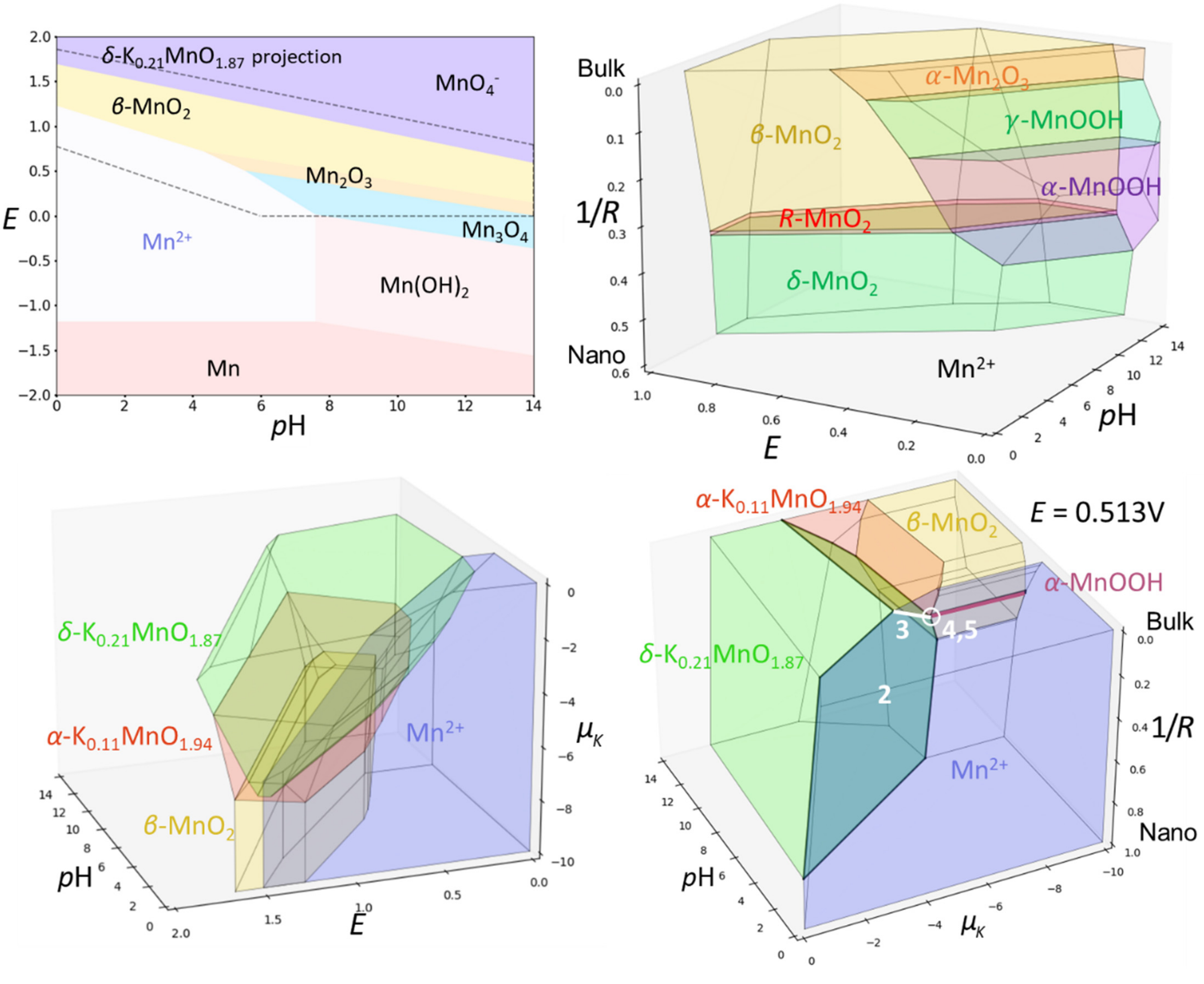

**Figure 5. Projections and slices of a high dimensional Pourbaix diagram in *p*H-*E*-$\mu_K$-1/*R* space of K-Mn-O-H system into lower dimensions.** (a) Traditional Pourbaix diagram with E and *p*H as axis. (b) A slice of *α*-$K_{0.11}MnO_{1.94}$, *δ*-$K_{0.21}MnO_{1.87}$, *β*-$MnO_2$, $Mn^{2+}$ when fixing *E* = 0.2 V. (c) A projection of *δ*-$K_{0.21}MnO_{1.87}$, *δ*-$MnO_2$, $Mn^{2+}$ in pH-*E*-$\mu_K$ space. (d) A slice of *α*-$K_{0.11}MnO_{1.94}$, *δ*-$K_{0.21}MnO_{1.87}$, *β*-$MnO_2$, $Mn^{2+}$ when fixing *E* = 1.2V. 2-phase coexistence regions are bounded by bold black lines.

However, adding surface energy contributions to the free energy of a material always reduces its acid stability. This is because surface energy is always positive, meaning that a high (1/R) increases the free-energy of a solid; whereas the free energy of the $Mn^{2+}$(aq) ion does not change with (1/R). From a Clausius-Clapeyron formulation, this is written as

$$\frac{\partial p\mathrm{H}}{\partial(1/R)} = \frac{\Delta(\gamma\eta\rho)}{\Delta RT\ln(10)(2N_O - N_H)}$$

Because $\gamma_{solid} > 0$ and $\gamma_{Mn2+(aq)} = 0$, the phase boundary as a function of acid stability always destabilizes the solid phase.

As defined by the Pourbaix potential, each single phase is a 4D polytope in the $p$H-$E$-1/$R$-$\mu_K$ space. Within this framework, a 2-phase coexistence region, designated as $\alpha$ and $\beta$ phase, is characterized by the condition $\Psi_\alpha = \Psi_\beta$, or equivalently, $\Delta\Psi_{\alpha-\beta} = 0$. **Table 4** shows for each intensive variable the difference in their conjugate extensive variables; for example, for $p$H this term would be $\Delta[-RT\ln(10)(2N_O\text{-}N_H)]$. From these coefficients, one can directly calculate the partial derivatives between any two natural variables, offering quantitative insight into how to affect relative stability. An in-depth discussion on how to represent coexistence from 2 to 5 phases using the Cartesian form of the Clausius-Clapeyron relations is presented in **Supplemental Information 2**.

**Table 4**. Coefficient vectors of 2-phase coexistence among $Mn^{2+}$ and $\beta$-$MnO_2$, $\alpha$-$K_{0.11}MnO_{1.94}$, $\delta$-$K_{0.21}MnO_{1.87}$, α-MnOOH

| Intensive variable | $p$H | $\mu_K$ | $E$ | 1/$R$ | - |
|---|---|---|---|---|---|
| Conjugated extensive quantity | $\Delta[-RT\ln(10)(2N_O\text{-}N_H)]$ | $-\Delta N_K$ | $\Delta[-(2N_O\text{-}N_H+Q)]$ | $\Delta(\gamma\eta\rho)$ | $\Delta(G_{bulk}\text{-}N_O\mu_{H2O})$ |
| Units | eV/Mn | eV/$K_{atom}$/Mn | eV/V/Mn | eV·nm/Mn | eV/Mn |
| $\beta$-$MnO_2$ + $Mn^{2+}$ | 0.105 | 0 | 0.886 | -0.451 | -1.090 |
| $\alpha$-$K_{0.11}MnO_{1.94}$ + $Mn^{2+}$ | 0.099 | 0.047 | 0.809 | -0.578 | -0.659 |
| $\delta$-$K_{0.21}MnO_{1.87}$ + $Mn^{2+}$ | 0.122 | 0.116 | 0.963 | -0.210 | -0.520 |
| $\alpha$-MnOOH + $Mn^{2+}$ | 0.125 | 0 | 0.707 | -0.696 | -1.139 |

On the other hand, when increasing the aqueous [$K^+$] concentration in the system—which increases $\mu_K$ by $\mu_K = \mu^0_K + RT\ln([K^+])$—the phases cryptomelane, $\alpha$-$K_{0.11}MnO_{1.94}$, and birnessite $\delta$-$K_{0.21}MnO_{1.87}$ appear on the phase diagram, with a much larger stability region relative to pyrolusite β-$MnO_2$. This can be rationalized by the open crystal structures of α-$MnO_2$ and δ-$MnO_2$ phases, as α-

$MnO_2$ Hollandite has 2×2 tunnel structures where intercalation of large K+ ions is energetically favorable, and similarly the δ-$MnO_2$ phase is a layered phase that also readily uptakes $K^+$. Intercalation of $K^+$ ions therefore stabilizes and lowers the bulk free energy of α- and δ- $MnO_2$, enlarging their stability windows. From a Clausius-Clapeyron formulation, this is written as

$$\frac{\partial pH}{\partial \mu_K} = -\frac{\Delta(N_K)}{\Delta RT \ln(10)(2N_O - N_H)}$$

The relative stability regions are visualized in the $p$H-$E$-$\mu_K$ space in **Figure 5c**. Notably, $\delta$-$K_{0.21}MnO_{1.87}$ is stable within a relatively large redox potential window at low $p$H.

As an aside; if we were performing experiments to solve for the Clausius-Clapeyron relations, even if our ultimate goal were to understand the derivatives d$p$H/d$Y$, we do not strictly need to do this measurement. As discussed previously, by choosing 3 of the 6 possible ratios: $d$$p$H/$dE$, $d$$p$H/$d$(1/R), $d$$p$H/$d\mu_K$, $dE$/$d$(1/R), $dE/d\mu_K$, or $d$(1/R)/$d\mu_K$—so long as all 4 variables are included—we can solve for the other three ratios.

## Engineering relative stability in four dimensions

With the full four-dimensional Clausius-Clapeyron relation from **Table 4**, we can make holistic assessments on how to engineer relative materials stability along four thermodynamic axes dimensions. From the perspective of an acid-stable manganese oxide catalyst, there are two primary considerations: First, the material should be the equilibrium phase under operation conditions, which can be affected by all four variables E, $p$H, 1/R and $\mu_K$. The second design consideration is from the perspective of acid-stability, where the phase boundary between the solid compound and $Mn^{2+}$ should have a d$p$H/$d$Y as negative as possible, for all considered intensive variables.

From these considerations, and from the manganese oxides in our dataset, the best acid-stable manganese oxide phase should be birnessite δ-$K_xMnO_2$. Of all the possible manganese oxide phases, δ-$MnO_2$ has the lowest surface energy due to its easily exfoliable 2D layered structure. This low surface energy means that the acid stability of δ-$MnO_2$ is least affected when increasing the surface area to volume ratio, for example by making low-dimensional nanoscale catalysts. From a functional perspective, catalysts rely on high surface area to volume ratios to maximize catalytic area, and δ-$MnO_2$ solubility increases the least at the nanoscale, compared to the other candidate $MnO_x$ phases.

Moreover, under a high aqueous concentration of $K^+$ ions, both α-$MnO_2$ and δ-$MnO_2$ can favorably intercalate K+ and reduce their bulk free-energies, which further increases their acid stability. Although d$p$H/d$\mu_{K+}$ is higher for α-$MnO_2$ than δ-$MnO_2$, α-$MnO_2$ is only stable in a narrow [$K^+$] concentration, whereas at higher [$K^+$] concentrations δ-$MnO_2$ is the stable phase. Therefore, it is more robust to simply use a high excess concentration of [$K^+$] to stabilize the δ-$MnO_2$ polymorph, which increases the reliability and operational stability of this functional compound.

## Three-, Four- and Five-Phase Coexistence

Gibbs' prediction of the triple-point on the temperature-pressure phase diagram; where solid, liquid and gas all coexist simultaneously; played a historical role in the adoption and establishment of chemical thermodynamics[45,46]. On higher-dimensional phase diagrams, three-phase coexistence is not only common, but an even greater number of possible coexisting phases is possible.

Designing conditions for multi-phase coexistence is promising in various functional devices. For example, photoelectrochemical water splitting[47], solid-state lithium-ion batteries[48], core-shell nanoparticle architectures[49] all rely on composite materials with hierarchical structures where there are multiple phases interacting with one another. Finding thermodynamic conditions for multiphase heterogeneous equilibrium can establish the long-term operational stability of such devices, which may otherwise degrade by undesired chemical reactions. In technologies that rely on phase transitions, such as multiferroric switching materials for transducers and information storage, finding a high-dimensional phase coexistence point may enable switching between more than two-states,[50] which could result in exciting new materials functionality.[51]

To illustrate 3- and 4-phase coexistence, we begin by analyzing **Figure 5d** as a representative 3D phase diagram. Here, with a fixed redox potential, Gibbs' Phase Rule is effectively the same as a diagram with just 3 intensive axes. In this case, two-phase coexistence is represented by 2D planes. The intersection of two planes (say, $\alpha+\beta$ and $\beta+\gamma$) produces a three-phase coexistence line ($\alpha + \beta + \gamma$). The intersection of two three-phase coexistence lines then leads to a 4-phase coexistence point.

However, **Figure 5d** is not a 3D phase diagram, but rather, is a 3D slice of a 4D phase diagram at a fixed redox potential. When we vary the value of the redox potential slice, the 4-phase coexistence point moves in the $p$H, 1/R and $\mu_K$ directions; this represents a 1D line on the 4D phase diagram. An animation of the 4D phase diagram is provided in Supplementary Movie 1, where we use time to illustrate the fourth dimension. There is one special point in 4D space, where the 4D line terminates in conjunction with another 4D coexistence line. At this specific vertex, there is 5-phase coexistence. We can in fact visualize this special point by finding the precise redox potential where this 5-phase coexistence happens, which is at $E$ = 0.513 V, which is the condition illustrated in Figure 4d. Therefore, the visualized 4D phase coexistence point on **Figure 5d** is in fact also a 5-phase coexistence point.

In **Table 5**, we provide explicit conditions for this 3-, 4- and 5-phase coexistence. These conditions are represented as vertices in 4D space. The 2-dimensional 3-phase coexistence boundary is fully defined by 3 vertices; and is spanned by any two of the 1D vectors that connect these 3 vertices. Similarly, the 1-dimensional 4-phase coexistence line is given by 2 vertices, and the 5-phase coexistence point is simply provided by its coordinates in 4D space.

**Table 5.** Three vertices of 3-phase coexistence among $\alpha$-$K_{0.11}MnO_{1.94}$, $\delta$-$K_{0.21}MnO_{1.87}$ and $Mn^{2+}$. Two vertices of 4-phase coexistence among $\beta$-$MnO_2$, $\alpha$-$K_{0.11}MnO_{1.94}$, $\delta$-$K_{0.21}MnO_{1.87}$ and $Mn^{2+}$. One vertex of 5-phase coexistence among $\beta$-$MnO_2$, $\alpha$-$K_{0.11}MnO_{1.94}$,$\delta$-$K_{0.21}MnO_{1.87}$, $Mn^{2+}$ and $\gamma$-MnOOH.

| | ***p*H** | $\mu_K$ | ***E*** | **1/*R*** |
|---|---|---|---|---|
| Three-phase coexistence $Mn^{2+}$ +$\alpha$-$K_{0.11}MnO_{1.94}$ + $\delta$-$K_{0.21}MnO_{1.87}$ | 0 | 4.353 | -1.120 | 0 |
| | 6.801 | -4.906 | 0.323 | 0 |
| | 7.368 | -6.763 | 0.513 | 0.211 |
| Four-phase coexistence $\beta$-$MnO_2$ + $\alpha$-$K_{0.11}MnO_{1.94}$ + $\delta$-$K_{0.21}MnO_{1.87}$ + $Mn^{2+}$ | -1.394 | -6.170 | 1.513 | 0.234 |
| | 6.680 | -6.732 | 0.553 | 0.222 |
| Five-phase coexistence $\beta$-$MnO_2$ + $\alpha$-$K_{0.11}MnO_{1.94}$ + $\delta$-$K_{0.21}MnO_{1.87}$ + $Mn^{2+}$ + $\alpha$-MnOOH | 7.368 | -6.763 | 0.513 | 0.211 |

## Conclusions

In Part I and Part II of this three-part series, we presented methods to construct high-dimensional phase diagrams, first with axes of extensive variables in Part I, and then with axes of intensive variables in Part II. However, since high-dimensional objects are so far removed from our everyday experience, these phase diagram can be difficult to navigate even when they are constructed. It can also be laborious and expensive to build high-dimensional phase diagrams in full, as for each axis we need all the thermochemical properties of each phase.

Our goal in Part III was to explore more the properties of phase boundaries, which again, are the key geometric objects on a phase diagram. Importantly, the gradient of a phase boundary is enough to evaluate relative stability, such that one does not necessarily need to characterize all the thermochemical data in a system. This transforms us from a 'thermodynamic assessment' process, where we construct the full phase diagram at once, to a more flexible framework that is quicker to implement in real-world engineering situations. One only has to characterize the experimentally-obtained phase, and then use concepts of relative stability to shift the applied experimental conditions towards the direction of the phase we desire. As more undesired phases are observed, one can iteratively build towards a full description of the high-dimensional phase boundaries between a target phase and all its competing phases. This offers a practical (and practicable) pathway to optimize the synthesis or operation conditions of target functional materials.

We conclude this three-part series by re-iterating Gibbs' first sentence in thermodynamics: "*Although geometrical representations of propositions in the thermodynamics of fluids are in general use, and have done good service in disseminating clear notions in this science, yet they have by no means received the extension in respect to variety and generality of which they are capable*." Despite a rich 150-year-old history, equilibrium thermodynamics still has many exciting opportunities for fundamental development.

Altogether, this three-part series provides a foundation to construct, navigate, and interpret new varieties of phase diagrams, with thermodynamic axes beyond temperature, pressure and composition, with as many axes as needed to capture all the essential physics of the thermodynamic system. As is often the case with thermodynamics, it is not strictly necessary to have perfect thermochemical data before we can derive meaningful scientific insights, or formulate promising engineering decisions. Most importantly, one needs a robust understanding of the geometric *structure* of thermodynamics. If one can visualize and anticipate the underlying geometry of free energy surfaces, as well as the conditions of heterogeneous equilibrium, we will altogether eliminate confusion and strengthen our intuition for materials design. Hopefully, this will enable us to better exercise our creativity to design the complex functional materials that drive modern technology.

## Code Availability

All code for analyzing and visualizing high dimensional Pourbaix diagrams and Clausius Clapeyron relations can be found on Github at the following link:

https://github.com/dd-debug/chemical_potential_diagram_and_convex_hull_and_pourbaix_diagram

The link includes a readme, tutorial example files, installation guide, Python package requirements, and instructions for use.


## Acknowledgements

Funding provided by the U.S. Department of Energy, Office of Science, Basic Energy Sciences, Division of Materials Science, through the Office of Science Funding Opportunity Announcement (FOA) Number DE-FOA-0002676: Chemical and Materials Sciences to Advance Clean-Energy Technologies and Transform Manufacturing.

[41] Birkner, Nancy, and Alexandra Navrotsky. "Rapidly reversible redox transformation in nanophase manganese oxides at room temperature triggered by changes in hydration." *Proceedings of the National Academy of Sciences* 111.17 (2014): 6209-6214.

[42] Navrotsky, Alexandra, et al. "Nanophase transition metal oxides show large thermodynamically driven shifts in oxidation-reduction equilibria." *Science* 330.6001 (2010): 199-201.

[43] Birkner, Nancy, and Alexandra Navrotsky. "Thermodynamics of manganese oxides: Sodium, potassium, and calcium birnessite and cryptomelane." *Proceedings of the National Academy of Sciences* 114.7 (2017): E1046-E1053.

[44] Birkner, Nancy, and Alexandra Navrotsky. "Thermodynamics of manganese oxides: Effects of particle size and hydration on oxidation-reduction equilibria among hausmannite, bixbyite, and pyrolusite." *American Mineralogist* 97.8-9 (2012): 1291-1298.

[45] Van Klooster, H. S. "Bakhuis Roozeboom and the phase rule." *Journal of Chemical Education* 31.11 (1954): 594.

[46] Caldi, Daniel G., and George D. Mostow, eds. *Proceedings of the Gibbs Symposium: Yale University, May 15-17, 1989*. American Mathematical Soc., 1990.

[47] Liu, Jiao, et al. "Surface-phase junctions of branched TiO2 nanorod arrays for efficient photoelectrochemical water splitting." *Applied Catalysis B: Environmental* 158 (2014): 296-300.

[48] Richards, William D., et al. "Interface stability in solid-state batteries." *Chemistry of Materials* 28.1 (2016): 266-273.

[49] Chu, Fukai, et al. "Hierarchical core–shell TiO2@ LDH@ Ni (OH) 2 architecture with regularly-oriented nanocatalyst shells: Towards improving the mechanical performance, flame retardancy and toxic smoke suppression of unsaturated polyester resin." *Chemical Engineering Journal* 405 (2021): 126650.

[50] Liu, Z. K., Xinyu Li, and Q. M. Zhang. "Maximizing the number of coexisting phases near invariant critical points for giant electrocaloric and electromechanical responses in ferroelectrics." *Applied Physics Letters* 101.8 (2012): 082904.

[51] Wei, Yanping, et al. "Four-state memory based on a giant and non-volatile converse magnetoelectric effect in FeAl/PIN-PMN-PT structure." *Scientific Reports* 6.1 (2016): 30002.

# The geometry of high-dimensional phase diagrams:
# III. Engineering relative stability in four dimensions

Jiadong Chen[1], Matt Powell-Palm[2], Wenhao Sun[1]*

[1]Department of Materials Science and Engineering, University of Michigan, Ann Arbor, Michigan, 48109, United States

[2]Department of Mechanical Engineering, Texas A&M University, College Station, TX, 77802, USA

*Correspondence to: whsun@umich.edu

# Supplemental Information

Contents

## SI1. Inferring missing thermochemical data with the Clausius-Clapeyron Relation

Gathering thermochemical data, either by computation or experiment, is always costly. However, we can often infer missing thermochemical data so long as one has enough information to fully constrain all relevant free-energy surfaces. For example, taking the *T-P-H* phase diagram of manuscript **Figure 3**, if we have already measured $\partial H/\partial P$ and $\partial T/\partial P$ then we can directly infer $\partial H/\partial T$. Note that for clarity we have omitted the full mathematical notation, but each of these partial derivatives is evaluated with all other variables held constant.

An exciting result emerges for two-phase coexistence in higher dimensions, where the amount of relative information needed to fully define the gradient of a high-dimensional phase boundary diminishes combinatorically. For any *k*-dimensional phase boundary, *k* number of 1D Clausius-Clapeyron vectors are needed to span the phase boundary. However, the number of possible $\partial Y_1/\partial Y_2$ relations increases as ${}_dC_2$. For example, on a 4D phase diagram, the phase boundary for 2-phase coexistence is 3-dimensional, and requires 3 spanning Clausius-Clapeyron vectors. However, there are 6 combinations of $\partial Y_1/\partial Y_2$ that can be measured {A/B, A/C, A/D, B/C, B/D, C/D}, and the measurement of any 3 ratios that contain all 4 variables constrains the entire phase boundary. **Table S2** extends this idea to higher dimensions, and shows the somewhat counterintuitive result that the higher dimensional the phase boundary, the fewer of the available individual Clausius-Clapeyron relations that need be measured to fully define the two-phase boundary. This approach is especially valuable when some of the partial derivatives are difficult to experimentally measure for some reason or another.

**Table S2:** Relative decrease in number of Clausius-Clapeyron measurements needed to define a two-phase phase boundary

| Dimension of Phase Boundary | Minimum $\partial Y_1/\partial Y_2$ measurements needed | Possible $\partial Y_1/\partial Y_2$ measurements | Fraction of required derivatives |
|---|---|---|---|
| 1 | 1 | 1 | 100% |
| 2 | 2 | 3 | 66% |
| 3 | 3 | 6 | 50% |
| 4 | 4 | 10 | 40% |
| 5 | 5 | 15 | 33% |
| 6 | 6 | 21 | 29% |

Not only can the generalized Clausius-Clapeyron relation be used to calculate high-dimensional phase boundaries from extensive material properties—it can also be used to extract difficult-to-measure extensive material properties *from* measured high-dimensional phase boundaries. However, it is important to note that even if we have the gradient of the plane and the molar extensive variables of one phase, we cannot directly infer *all* of the molar extensive variables of the other phase simultaneously. This can be shown from a matrix equation **Ax** = **B**:

$$\begin{bmatrix} \partial H / \partial P & 0 & -1 \\ 0 & \partial T / \partial P & -1 \\ \partial H / \partial T & -1 & 0 \end{bmatrix} \begin{bmatrix} M_\gamma \\ S_\gamma \\ V_\gamma \end{bmatrix} = \begin{bmatrix} (\partial H / \partial P) M_\alpha - V_\alpha \\ (\partial T / \partial P) S_\alpha - V_\alpha \\ (\partial H / \partial T) M_\alpha - S_\alpha \end{bmatrix} \qquad \textbf{Eq. 7}$$

This matrix equation has a unique solution if and only if rank(**A**) = $n$, where $n$ is number of elements in **x**. If, on the other hand, the determinant of **A** is zero, then **A** is singular and there are linearly dependent columns of matrix **A**. The determinant of matrix **A** in **Eq. 7** is 0, therefore **Eq. 7** does not have a unique solution for $S_\gamma$, $V_\gamma$, $M_\gamma$.

Therefore, to calculate the molar extensive quantities of the $\gamma$ phase, we need (at least) one initial value from any of $S_\gamma$, $V_\gamma$, $M_\gamma$; after which the system is fully constrained and we can calculate the remaining two molar extensive quantities. This rank analysis can be generally extended to high-dimensions as well.

Consistent with the long-standing tradition of experimental classical thermodynamics, this generalized Clausius-Clapeyron formulation thus presents myriad opportunities to leverage combinations of readily measurable extensive and intensive thermodynamic data to infer the values of less-accessible quantities of both categories. Future work may explore the combinatorial limits of these options, to great potential value to the modern experimentalist.

## SI2. Examples of Cartesian representation of Clausius-Clapeyron relationships

The Cartesian representation of the generalized Clausius-Clapeyron relationship helps determine how one natural variable will change in relation to others in situations of multi-phase coexistence and in higher dimensions. In our manuscript, we demonstrate the coexistence of 1 to 5 phases within the $p$H-$E$-1/$R$-$\mu_K$ 4D Pourbaix diagram. All calculations are conducted using this Cartesian form. One of the key advantages of employing the Clausius-Clapeyron equation in this form is our ability to precisely calculate the dynamic relationships between different natural variables. By utilizing the materials' formation energies and extensive molar quantities, we are equipped to compute changes in an intensive natural variable relative to others under varying conditions.

### SI2.1. Single phase

As explored in Part II of this three-part series on high-dimensional phase diagram papers, each single phase is characterized through a Legendre transformation, which is resolved by the intersection of half-spaces. In the manuscript, we present the Pourbaix potential ($\Psi$) for each single phase, in relation to the four natural variables: $p$H, electrode potential ($E$), inverse temperature (1/$R$), and the chemical potential of potassium ($\mu_K$). Within the $p$H-$E$-1/$R$-$\mu_K$ space, each phase is represented as a four-dimensional (4D) polytope:

$$\Psi = \frac{1}{N_m}\left( (G_{bulk} - N_O\mu_{H_2O}) - RT\ln(10)(2N_O - N_H)p\mathrm{H} - (2N_O - N_H + Q)E + (\frac{A}{V})\gamma\eta\rho - N_K\mu_K \right)$$

### SI2.2. 2-phase coexistence

A 2-phase coexistence region in a 3D space is a 2D surface shared by two 3D polytopes, as shown in the chemical potential diagram of **Figure S2**. Similarly, a 2-phase coexistence region in a 4D space is a 3D interface shared by two 4D polytopes. Just as we are accustomed to presenting the equation of a 2D plane ($ax + by + cz + d = 0$) within a 3D space (defined by the coordinates $x$, $y$, and $z$), we can represent a 3D 2-phase coexistence region within a 4D space through a linear combination of the four natural variables:

$$[X_{p\mathrm{H}}]_{\alpha-\beta} \cdot p\mathrm{H} + [X_{\mu_K}]_{\alpha-\beta} \cdot \mu_K + [X_E]_{\alpha-\beta} \cdot E + [X_{1/R}]_{\alpha-\beta} \cdot \frac{1}{R} + [C]_{\alpha-\beta} = 0 \qquad \textbf{Eq. 1}$$

where $[X_{p\mathrm{H}}]_{\alpha\text{-}\beta}$, $[X_{\mu K}]_{\alpha\text{-}\beta}$, $[X_E]_{\alpha\text{-}\beta}$, $[X_{1/R}]_{\alpha\text{-}\beta}$, $[C]_{\alpha\text{-}\beta}$ represent coefficient vectors – the conjugated variables of the four natural variables and a constant, their physical form is represented in manuscript **Table 4**. This approach simplifies the complex task of delineating phase coexistence in higher-dimensional spaces, making it more accessible for analysis and interpretation.

Thus, a vector of coefficients $[X_{p\mathrm{H}}]_{\alpha\text{-}\beta}$, $[X_{\mu K}]_{\alpha\text{-}\beta}$, $[X_E]_{\alpha\text{-}\beta}$, $[X_{1/R}]_{\alpha\text{-}\beta}$, $[C]_{\alpha\text{-}\beta}]$ can represent a 3D 2-phase coexistence in the 4D $p$H-$E$-1/$R$-$\mu_K$ space. Their values can be calculated by employing the condition $\Psi_\alpha = \Psi_\beta$ at the $\alpha$-$\beta$ phase coexistence. Consequently, the partial derivative between any two natural variables can be expressed as the ratio of the negative reciprocal of the corresponding two coefficients, such as $\partial\mu_K / \partial p\mathrm{H} = -[X_{p\mathrm{H}}]_{\alpha-\beta} / [X_{\mu_K}]_{\alpha-\beta}$.

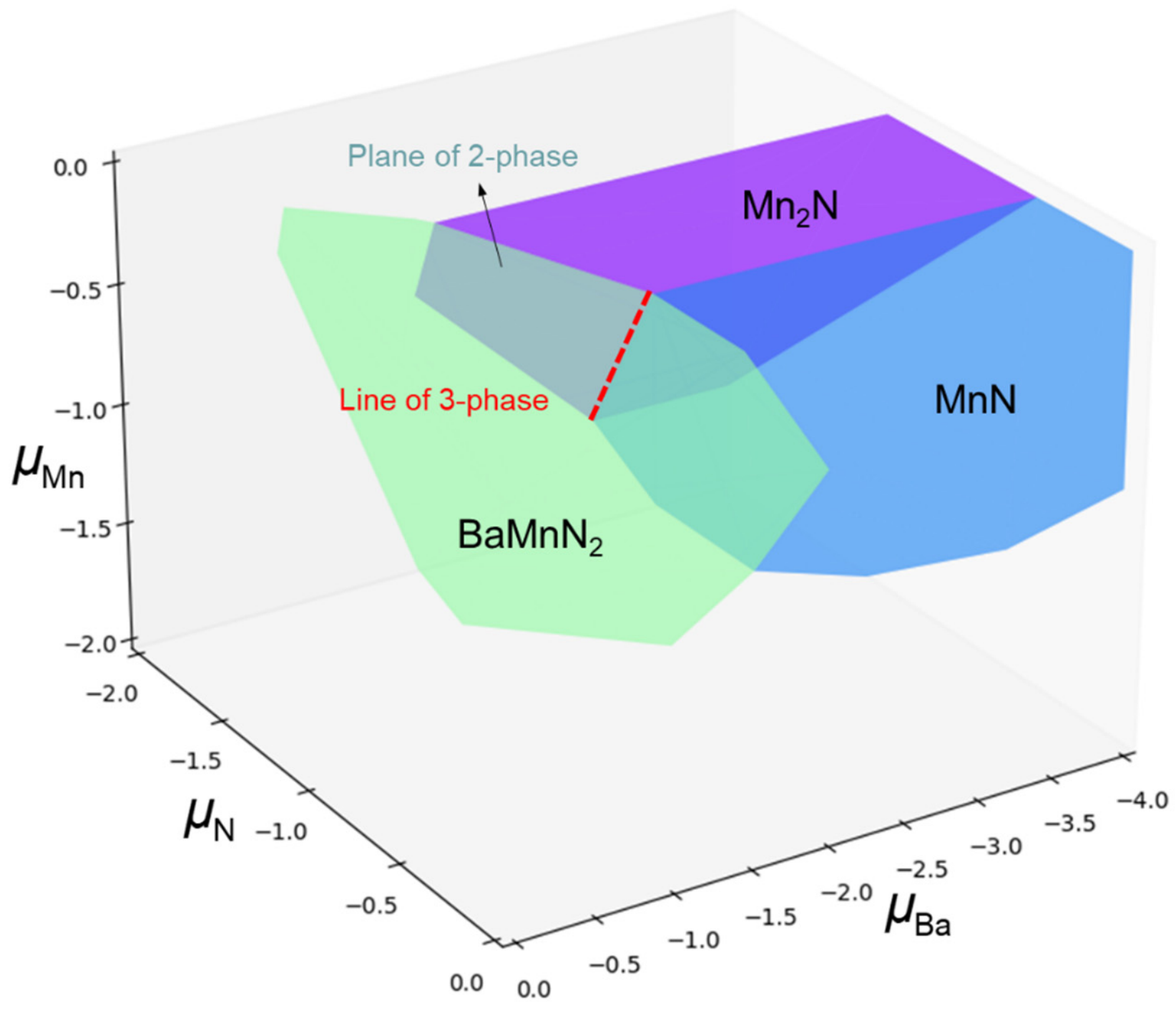


**Figure S1**. Chemical potential diagram of $BaMnN_2$, $Mn_2N$ and MnN. The red dashed line represents the 3-phase coexistence region. The shared planes by two single phases are the 2-phase coexistence regions.

**SI2.2. 3-phase coexistence**

For a ternary system in a 3D space, a 3-phase coexistence region is 1D, as shown as the red dashed line in **Figure S1**. Mathematically, a 2D 2-phase coexistence plane in this 3D space is A·$\mu_{Ba}$ + B·$\mu_{Mn}$ + C·$\mu_N$ + D = 0. We can use the Eq.2 and Eq.3 to represent the phase coexistence between $BaMnN_2$ + $Mn_2N$ and MnN + $Mn_2N$, respectively.

$$[X_{\mu Ba}]_1 \cdot \mu_{Ba} + [X_{\mu Mn}]_1 \cdot \mu_{Mn} + [X_{\mu N}]_1 \cdot \mu_N + [C]_1 = 0 \quad \textbf{Eq. 2}$$

$$[X_{\mu Ba}]_2 \cdot \mu_{Ba} + [X_{\mu Mn}]_2 \cdot \mu_{Mn} + [X_{\mu N}]_2 \cdot \mu_N + [C]_2 = 0 \quad \textbf{Eq. 3}$$

A 1D 3-phase coexistence line, is the intersection of two 2D 2-phase coexistence planes. So, the $BaMnN_2$ + $Mn_2N$ + MnN 3-phase coexistence line must simultaneously satisfy Eq.2 and Eq.3 simultaneously. Here, any variable can be eliminated by forming a linear combination of Eq.2 and Eq.3. For instance, by multiplying Eq.3 by -$[X_{\mu N}]_1/[X_{\mu N}]_2$ and adding it to Eq.2, we establish a linear relationship between $\mu_{Ba}$ and $\mu_{Mn}$. Therefore, a line in a 3D space indicates that any two among the three natural variables share a linear relationship.

Similar statements are also valid for the 4D $p$H-$E$-1/$R$-$\mu_K$ space. Just like a 3-phase coexistence region in a 3D space is a 1D line; a 3-phase coexistence region in a 4D space is a 2D plane – the intersection of two 3D 2-phase coexistence polytopes (Eq.4 and Eq.5).

$$[X_{p\mathrm{H}}]_1 \cdot p\mathrm{H} + [X_{\mu_K}]_1 \cdot \mu_K + [X_E]_1 \cdot E + [X_{1/R}]_1 \cdot \frac{1}{R} + [C]_1 = 0 \quad \textbf{Eq. 4}$$

$$[X_{p\mathrm{H}}]_2 \cdot p\mathrm{H} + [X_{\mu_K}]_2 \cdot \mu_K + [X_E]_2 \cdot E + [X_{1/R}]_2 \cdot \frac{1}{R} + [C]_2 = 0 \quad \textbf{Eq. 5}$$

Again, any variable can be eliminated by linear combination of Eq.4 and Eq.5.Therefore, a 2D plane in a 4D space means any three among four natural variables has a linear relationship, which can be expressed by:

$$\begin{aligned}
&[X_{p\mathrm{H}}]_1 \cdot p\mathrm{H} + [X_{\mu_K}]_1 \cdot \mu_K + [X_E]_1 \cdot E + [C]_1 = 0 \\
&[X_{p\mathrm{H}}]_2 \cdot p\mathrm{H} + [X_{\mu_K}]_2 \cdot \mu_K + [X_{1/R}]_2 \cdot {}^1\!/\!_R + [C]_2 = 0 \\
&[X_{p\mathrm{H}}]_3 \cdot p\mathrm{H} + [X_{1/R}]_3 \cdot {}^1\!/\!_R + [X_E]_3 \cdot E + [C]_3 = 0 \\
&[X_{1/R}]_4 \cdot {}^1\!/\!_R + [X_{\mu_K}]_4 \cdot \mu_K + [X_E]_4 \cdot E + [C]_4 = 0
\end{aligned} \quad \textbf{Eq. 6}$$

Eq. 6 gives us four sets of coefficient vectors, where an example for $Mn^{2+}$+α-$K_{0.11}MnO_{1.94}$+δ-$K_{0.21}MnO_{1.87}$ 3-phase coexistence is shown in manuscript **Table S1**. From a Clausius-Clapeyron perspective, the four sets of coefficients can help us determine the change of an intensive variable as the change of another intensive variable under different fixed conditions. For example, under a fixed redox potential $E$, the relationship of pH and $\mu_K$ is $\partial \mu_K / \partial p\mathrm{H}|_E = -[X_{p\mathrm{H}}]_1 / [X_{\mu_K}]_1$ by Eq 6.1. And if particle radius R is assumed fixed, $\partial \mu_K / \partial p\mathrm{H}|_R = -[X_{p\mathrm{H}}]_2 / [X_{\mu_K}]_2$ by Eq 6.2. And if $R$ and $E$ are both fixed, based on Eq 6.3 and 6.4 $p$H and $\mu_K$ are fixed too. This means there are only two degrees of freedom for a 3-phase coexistence in a 4D phase diagram.

**Table S1.** Coefficient vectors of 3-phase coexistence among $Mn^{2+}$, *α*-$K_{0.11}MnO_{1.94}$ and δ-$K_{0.21}MnO_{1.87}$.

| | A | B | C | D | F |
|---|---|---|---|---|---|
| $Mn^{2+}$ + α-$K_{0.11}MnO_{1.94}$ + δ-$K_{0.21}MnO_{1.87}$ | 0 | 0.058 | -0.040 | 0.506 | 0.297 |
| | -0.120 | 0 | -1.023 | 1.209 | 1.099 |
| | 0.005 | 0.060 | 0 | 0.478 | 0.265 |
| | 0.086 | 0.099 | 0.669 | 0 | -0.280 |

### SI2.3. 4-phase and 5-phase coexistence

Similarly, a 4-phase (*α*, *β*, *γ*, *δ*) coexistence region is the intersection of three 3D 2-phase (*α*-*β*, *β*-*γ*, *γ*-*δ*) coexistence regions, which is a line. So, any two variables can be eliminated by linear combination of three 2-phase equations (Eq.1). Therefore, a 1D line in a 4D space means any two among four natural variables has a linear relationship. This requires $^{2}C_{4}$=6 equations to represent this line. But we only need two vertices to represent this 4-phase coexistence line by using parametric form, as shown as manuscript Table 5.

For a 5-phase coexistence region, it is a 0D point. Any three variables can be eliminated by linear combination of four 2-phase equations (Eq.1). It means any natural variables has a fixed value (the coefficients of other variables are 0).

**Table S3.** Bulk Formation Energies, Surface Energies, Shape factors, Volume/Metal of K-Mn-O-H Phases

| **Phase** | Formation energy | Surface energy | Shape factor | Volume/metal |
|---|---|---|---|---|
| **Unit** | eV/ formula | $J/m^2$ | -- | $Å^3/Mn$ |
| R-$MnO_2$ | -4.783 | 1.33 | 3.53 | 29.7 |
| $\alpha$-$K_{0.0625}MnO_2$ | -5.03 | 1.19 | 5.35 | 33.8 |
| $\alpha$-$K_{0.11}MnO_{1.94}$ | -5.6 | 1.19 | 5.35 | 33.8 |
| $\alpha$-$K_{0.125}MnO_2$ | -5.364 | 1.19 | 5.35 | 33.8 |
| $\alpha$-$K_{0.166}MnO_2$ | -5.52 | 1.19 | 5.35 | 33.8 |
| $\alpha$-$K_{0.25}MnO_2$ | -5.764 | 1.19 | 5.35 | 33.8 |
| $\alpha$-$Mn_2O_3$ | -9.132 | 1.19 | 5.35 | 33.8 |
| $\alpha$-$MnO_2$ | -4.767 | 1.19 | 5.35 | 33.8 |
| $\alpha$-MnOOH | -5.763 | 1.19 | 5.35 | 33.8 |
| $\beta$-$MnO_2$ | -4.837 | 1.54 | 3.85 | 27.5 |
| $\beta$-MnOOH | -5.629 | 1.54 | 3.85 | 27.5 |
| $\delta$-$K_{0.21}MnO_{1.87}$ | -6.02 | 0.14 | 9.79 | 44.4 |
| $\delta$-$K_{0.33}MnO_2$ | -5.988 | 0.14 | 9.79 | 44.4 |
| $\delta$-$K_{0.5}MnO_2$ | -6.469 | 0.14 | 9.79 | 44.4 |
| $\delta$-$K_{0.75}MnO_2$ | -6.894 | 0.14 | 9.79 | 44.4 |
| $\delta$-$MnO_2$ | -4.558 | 0.14 | 9.79 | 44.4 |
| $\gamma$-MnOOH | -5.964 | 0.84 | 6.09 | 33.5 |
| $\gamma$-$MnO_2$ | -4.787 | 0.84 | 6.09 | 33.5 |
| Mn | 0 | -- | -- | -- |
| MnO | -3.762 | -- | -- | -- |
| $Mn_2O_3$ | -9.132 | -- | -- | -- |
| $Mn_3O_4$ | -13.346 | 1.43 | 5.44 | 26.2 |
| $Mn(OH)_2$ | -6.198 | 0.47 | 5.69 | 43.5 |
| $KMnO_2$ | -7.313 | -- | -- | -- |
| $Mn^{2+}$ (aq) | -2.363 | -- | -- | -- |
| $MnO_4^-$ (aq) | -4.634 | -- | -- | -- |
| $K^+$ (aq) | -2.926 | -- | -- | -- |
| $Mn^{3+}$ (aq) | -0.85 | -- | -- | -- |
| $MnO_4^{2-}$ (aq) | -5.222 | -- | -- | -- |
| $Mn(OH)_3^-$ (aq) | -7.714 | -- | -- | -- |
| $MnOH^+$ (aq) | -4.198 | -- | -- | -- |
| $HMnO_2^-$ (aq) | -5.243 | -- | -- | -- |